\documentclass{emulateapj}

\usepackage{epsfig}
\usepackage{amsmath}
\usepackage{natbib}
\usepackage{chngpage}
\usepackage{multirow}
\usepackage{array}
\usepackage{graphicx}
\usepackage{epsfig, graphics}
\usepackage{bbm}
\fontsize{15}{20}
\long\def\symbolfootnote[#1]#2{\begingroup%
\def\thefootnote{\fnsymbol{footnote}}\footnote[#1]{#2}\endgroup} 

\begin{document}

\title{Chandra observations of AGN-candidates correlated with Auger UHECRs}

\author{ }
\affiliation{ }

\author{William A. Terrano}  
\affiliation{Department of Physics, University of Washington, Seattle, WA 98195, USA}
\author{Ingyin Zaw}
\affiliation{New York University Abu Dhabi, Abu Dhabi, UAE}
\author{Glennys R. Farrar}
\affiliation{Center for Cosmology and Particle Physics \&
Department of Physics\\ New York University, NY, NY 10003, USA}

\begin{abstract}
The Auger observatory has observed a possible correlation between Ultrahigh Energy Cosmic Rays (UHECRs) above 57 EeV and nearby candidate Active Galactic Nuclei (AGN) from the Veron-Cetty Veron catalog (VCV).  In this paper we report on \emph{Chandra} X-ray observations of 10 unconfirmed VCV AGN-candidates and luminous IR galaxies correlating with the first set of Auger UHECRs, to determine whether or not they have active nuclei.  The X-ray data, when combined with optical luminosities, show that in fact none of the 10 galaxies have a significant AGN component; if there is any nuclear activity at all, it is weak rather than obscured.  This reduces the number of UHECRs in the original Auger dataset possibly correlating with AGNs from 20 of 27 down to 14 of 27.   We also used \emph{Chandra} to measure the X-ray luminosity of \emph{ESO 139-G12}, an AGN which correlates with 2 of the Auger UHECRs, to obtain the first estimate of its bolometric luminosity; this completes the determination of the bolometric luminosities of all correlating AGNs.  Taking our results into account, only one of the 27 UHECRs in the original Auger data-release is correlated on a few-degree angular scale with an identified AGN that is powerful enough in its steady-state to accelerate protons to the observed energies, according to conventional acceleration mechanisms.  Intriguingly, $\approx 60$\% of the UHECRs with $|b|>10^\circ$ do correlate with genuine AGNs, but these are too weak to meet the acceleration criterion for protons; this may be an indication that AGNs experience transient high-luminosity states which can accelerate UHECRs.  To determine the source(s) and composition of UHECRs through statistical correlation studies requires reliable, complete and uniform catalogs of identified AGNs; our \emph {a posteriori} inspection of ambiguous source candidates underscores the inadequacies of the VCV catalog in this respect.


\end{abstract}
\maketitle

\section{Introduction}

Identifying the sources of ultra-high energy cosmic rays (UHECRs) is one of the major outstanding goals in astrophysics.  Progress has proven difficult due to the rarity of high energy events and because the deflections of the cosmic rays as they travel through intergalactic magnetic fields mean that the cosmic ray arrival directions do not point back to their origins. The very most energetic cosmic rays ($E \gtrsim 6\times10^{19}$ eV) have attracted attention as a promising path forward.  The energy loss due to the GZK mechanism means that CRs with such energies typically have traveled about 100 Mpc, significantly limiting the possible sources for such high energy particles. Furthermore, at these energies magnetic deflections of protons may be small enough to allow the identification of the progenitor type based on statistical associations.

By 2007, the Pierre Auger Collaboration (2007, 2008) had compiled enough events to begin drawing statistical correlations between UHECR events and possible sources, and reported a strong correlation between UHECRs and the nearby galaxies listed in the \citet{VCV} Catalog of Quasars and Active Galactic Nuclei ($12^{th}$ Ed.) \citep{augerScience07}.  The scan procedure that was used steps through UHECR energy threshold, maximum angular separation, and maximum VCV galaxy redshift to find the values giving the lowest chance probability compared to an isotropic distribution, then evaluates the likelihood of such a correlation occuring by chance by performing the same analysis on many isotropic datasets.  In the following, the term ``correlated" referring to a galaxy with respect to a UHECR simply means that the given galaxy falls within the angular and redshift limits returned by applying the scan method with that galaxy catalog.

Of the 27 cosmic rays with energies above 57 $\times 10^{18}$ eV that Auger detected before Aug. 31, 2007, twenty correlate within  $3.2^{\circ}$ with VCV galaxies having $z \leq 0.018$, about 75 Mpc \citep{augerLongAGN}.  Since more than one galaxy can be correlated with a UHECR and vice versa, there are 21 VCV galaxies correlated with these 20 UHECRs.  Galaxy catalogs such as VCV are incomplete in the Galactic Plane, so a better comparison is obtained by restricting to $|b|>10^{\circ}$ where the VCV catalog is more complete.  With this restriction there are 22 UHECRs of which 19 UHECRs correlate with 20 galaxies \citep{zfg09}.  This is a much higher correlation than would be expected by chance from an isotropic source distribution, and higher than can be explained by nearby galaxy clustering alone \citep{zbf10}. More recent Auger data continue to show a significant, albeit less strong, correlation with VCV galaxies \citep{augercollectedICRC09}.  

	
 An important question is whether the observed correlation with VCV galaxies implies that AGNs are the sources of some or all UHECRs. The VCV catalogue is a list of AGN \emph{candidates}, so first it must be established whether the correlating galaxies (i.e., the VCV galaxies within 3.2$^\circ$ of a UHECR) are actually AGNs or not.  \citet{zfg09} looked at existing observations of the galaxies in VCV that were correlated with the first set of UHECRs and found that only 14 of the 21 galaxies show unambiguous evidence of AGN activity.  Three show no signs of AGN activity, while the other four have ambiguous optical spectra. A widely used technique for optical identification of AGNs, so-called BPT line ratios, compares the relative strengths of diagnostic spectral lines \citep{bpt1981}.  The BPT line ratios of the 4 ambiguous galaxies fall within the ranges classified by \citet{Kauffmann2003} as AGNs but they do not fall within the line-ratio ranges adopted by \citet{KewleyRatio} as indicative of an active nucleus. AGN activity can often be obscured in the optical bands by dust obscuration, and the Kewley test excludes some known AGNs.  Additionally, as many as half of AGNs that are selected based on radio or X-ray properties would not be identified by looking only at their BPT line ratios \citep{reviglioHelfand06}.  Thus, determining whether these 4 ambiguous VCV galaxies have active nuclei requires observations outside of optical wavelengths.

The VCV galaxies are not the only population of nearby galaxies found to correlate with the Auger UHECRs. 
\citet{bzf10} find an anomalously large correlation between the Auger galaxies and nearby Luminous IR galaxies in the PSCz catalog.  Restricting to $|b|>10^\circ$ where PSCz is complete, 13 galaxies of $L_{IR} > 10^{10.5} L_{sun}$ correlate within $2.1^{\circ}$ with one or more of the 22 Auger UHECR events.  Some LIRGs contain an active nucleus with dust absorbing the AGN radiation and re-emitting it thermally, giving rise to large IR luminosities. This raises the question of whether the LIRGs found to correlate may actually be AGNs. \citet{bzf10} showed that 6 of the 13 correlating IR galaxies are also in VCV and 5 of these do in fact host AGNs\footnote{The one which is not a confirmed AGN falls within the VCV sample to be tested.}. The other 7 correlating IR galaxies lacked the observations needed to differentiate between star-formation and obscured nuclear activity.

A key distinguishing feature of active galaxies is that accretion-driven radiation produces a nearly flat spectrum from the IR to the X-ray, known as the broadband continuum.  Normal and starburst galaxies, on the other hand, have broadened blackbody spectra peaked sharply in the UV/optical. This means the X-ray to optical flux ratios are considerably harder for AGNs than for normal/starburst galaxies. If the active nucleus is obscured, the X-ray to optical ratio becomes even harder since dust absorbs UV and optical photons more readily than X-rays \citep{Comastri2003}. This makes X-ray observations, particularly when combined with observations in the near-IR, optical or UV, a powerful tool for identifying obscured AGNs. 

In this paper we use X-ray observations to determine whether the 4 ambiguous VCV galaxies, and the 7 indeterminate PSCz galaxies, have active nuclei.  We observed ten of these possible UHECR source galaxies using the \emph{Chandra} X-ray satellite, while for \emph{IC 5169} we used data from a recent XMM-Newton observation.

Another interesting question is whether AGNs that correlate with UHECRs have any characteristic features which distinguish them from the AGNs that do not seem to correlate with UHECRs.  A particularly relevant property is the luminosity of the AGN, as this helps to constrain the possible cosmic ray acceleration mechanisms \citep{FarrarandGruzinov}.  Previously, reliable luminosities have been established for all but one of the correlating AGNs \citep{zfg09}, \citep{bzf10}.  We observed the remaining AGN, \emph{ESO 139-G12},  in order to obtain the first robust estimate of its bolometric luminosity.

	\begin{table*}
	\begin{center} 
		\begin{tabular}{|l|c| p{1.8cm} | p{1.8cm} |p{1.8cm}|p{1.8cm}|c|c|m{1.5cm}|}
		\hline
		\tabletypesize{\scriptsize}
		Source Name & Obs ID & Total Counts \newline (0.5-10 keV) & Hard Counts \newline (2-10 keV) & Flux ~~ (Direct) & Flux (webPIMMS) & Rmag & $Log(f_x/f_R) $ &  $ L_{2-10}$ \\
		
		\hline
		IC 5169& 10265 & 17 & 6 & 3.3 E-14 & 1.4 E -14 & 12.5$^1$ & -3.0 & 7.4 E 39\\
		NGC 7591 & 10264 & 17 & 2 & 6.6 E -15 & 1.4 E - 14 & 12.4$^{3,b}$ & -3.4 & 8.6 E 39 \\
		NGC 1204 & 10256 & 13 & 4 & 1.3 E -14 & 1.2 E -14& 13.6 & -2.9 & 5.8 E 39\\
		NGC 2989 & 10255 & 5 & 2 &2.3 E -14 & 4.9 E -15 &  12.5$^{1,2}$ & -3.1 & 8.8 E 39\\
		\hline
		\hline
		IC 4523& 10261 & 6 & 5 & 6.5 E -14 & 6.5 E -15 & 13$^2$  & -2.5  & 3.8 E 40\\
		ESO 270-G007 & 10260 & 14 & 5 & 2.2 E -14 & -- & 13$^1$ & -3 & 8.4 E 39\\
		IC 5186 & 10259 & 5 & 0 & 5.8 E -15$^a$ & -- & 12.3$^1$ & -3.8  & 3.4 E 39\\
		ESO 565-G006 & 10258 & 21 & 8 & 5.5 E -14 & 2.4 E -14 & 12.7$^1$ & -2.7 & 3.2 E 40\\
		NGC 7648 & 10257 & 5 & 1 & 3.0 E -15 & -- & 12.2$^{3}$ & -4.4 & 9.7 E 38\\
		2MASX J1 754-60 & 10262 & 3 & 1 & 2.7 E -15 & -- & -- & -- & 1.6 E 39\\
		IC 5179 & XMM-Newton & 2685 & -- & -- & 9.1 E -14 & 11.4 & -2.98 & 2.5 E 40\\
		\hline
		\hline
		ESO 139-G12 & 10263 & 1070 & 666 & 4.7 E -12$^c$ & -- & 12.9$^{1,2}$ & -0.2 & 2.9 E 43\\
		\hline
		\end{tabular}
	\end{center}
	\caption{Table listing source properties}
		\label{SourceProp}
	\tablerefs{(1) \cite{ESOCAT1989}, (2) \cite{HIPASS}, (3) \cite{Vaucouleurs}}
	\tablecomments{Fluxes are for 2-10 keV and are in cgs units (erg cm$^{-2}$ s$^{-1}$); only the direct flux determination method is used for diffuse sources.   The 2-10 keV luminosity, $L_{2-10}$, is in units of erg s$^{-1}$ and is derived from the direct flux measurement; it is the total X-ray luminosity in the central region and thus an upper limit on X-ray luminosity of a possible AGN.  The first four rows are the correlating galaxies found in VCV.  The following six are those from PSCz. The next row was observed by XMM-Newton: the counts are from 0.5 - 12 keV and the observation lasted ~25000 secs.  The last row is a known AGN whose  X-ray luminosity had not previously been measured, needed to assess its ability to accelerate UHECRs.  \\ $(a)$ No counts above 2 keV. $(b)$ Johnson magnitude rather than Cousins. This does not affect the result. $(c)$ Derived from fit, see Figure 1.\\ }

\end{table*} 

\section{Observations and Data Reduction}

Each target was observed  on the Chandra ACIS-I detector for 5 kiloseconds between January and August 2009. Data analysis was performed using the {\tt ciao} data analysis software. All of the targets corresponded to an X-ray source. (Table \ref{SourceProp}).

For the \emph{Chandra} observations, we used two different methods to estimate the X-ray energy flux. The direct method sums the energy received per unit time in the band of interest (2-10 keV) in a 2 arcsec window around the source using the {\tt eff2evt} tool which takes into account both the quantum efficiency and the effective area corrections of the satellite. The flux is typically dominated by higher energy and less frequent events, making this technique vulnerable to Poisson fluctuations when the count rates are small.  The second technique considers only the total counts received from the compact nuclear source, ignoring the energies of the photons, and assumes that the spectrum follows a power law with a spectral index of 2. Then webPIMMS can estimate the corresponding flux, including Galactic extinction from webPIMMS and assuming no intrinsic extinction.  For the galaxies which are diffuse sources, we report only the direct measurement of the flux within 2 arcsec of the galactic center, as an upper limit on the possible nuclear emission. 

We also analyzed the XMM-Newton observation of \emph{IC 5179}.  Using the source imaging PPS processing provided by XMM, we took the count rate on the EPIC pn camera (0.5-12 keV) and used webPIMMS to estimate the corresponding 2-10 keV flux of the source.

\section{Analysis and Results}

X-ray to optical flux ratios provide a useful way to classify galaxies according to their nuclear activity. \citet{Barger2002} show that for AGNs, the ratio of the 2-10 keV flux to the R-band magnitude is typically $ 1 > log(f_x/f_R)$\symbolfootnote[2]{$log(f_x/f_R) \equiv log_{10}(f_x) + 5.5 + .4*R_{mag}$ where $f_x$ is the 2-10 keV flux}$ > -1$, while obscured AGNs have $log(f_x/f_R) > 1$.  Normal and starburst galaxies have  $log(f_x/f_R) < -2$, making this a robust way of discriminating AGNs, even obscured ones, from normal and starburst galaxies \citep{Donley2005}.  

The results are summarized in Table \ref{SourceProp}. The X-ray to optical ratios are far below what is expected from AGNs ($log(f_x/f_R) > -1$) for ten of the ambiguous or indeterminate galaxies: nine we observed with \emph{Chandra}, and \emph{IC 5179} for which we used an XMM-Newton observation.  This holds even when the larger of the two X-ray flux determinations is used to calculate the X-ray to optical flux ratios.  Since our measure of nuclear activity depends on $log_{10}(f_x)$, it is relatively insensitive to the uncertainties inherent in the determination of the flux at these low count rates.  We conclude that none of these galaxies have active nuclei. 

The last of the indeterminate galaxies (\emph{2MASX J1 754-60}) lacks the optical observations needed for the $log(f_x/f_R)$ test, but its X-ray emission is diffuse and weak, and its near-IR properties are similar to the other galaxies.  Therefore we conclude that it is unlikely to host an AGN, certainly not one expected to be capable of UHECR acceleration.

Obscured AGN in X-ray surveys are often identified by the hardness ratio of the source. The hardness ratio is defined as H-S/H+S where H is counts from 2-8 keV and S is counts from 0.5-2 keV. Typically, nearby AGN  with column densities greater than $N_H \approx 10^{22}$ have positive hardness ratios, while unobscured sources have negative hardness ratios \citep{Fiore2000}.  Table \ref{SourceProp} gives counts from  0.5-10 keV and 2-10 keV.  These can be used to estimate the hardness ratio since the small effective area of \emph{Chandra} at high energies meant that no photons with energies above 8 keV were detected for any of the sources.  Although the small number of counts means large uncertainties in the hardness ratios, they show no evidence of obscuration.

\begin{figure}
\includegraphics[width = 0.35 \textwidth,angle=270]{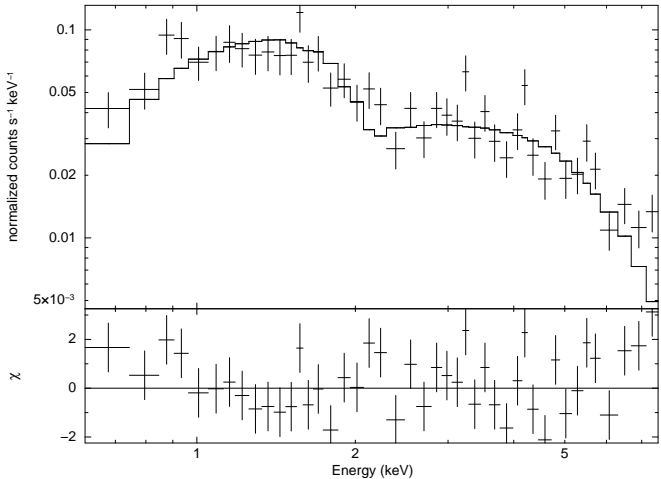}
\caption{Data and fit to a power law ($\Gamma = 0.72^{+0.09}_{-0.09}$, 90\% confidence) with N$_{\rm H} < 8.2 \times 10^{21} {\rm cm}^{-2}$ at 90\% confidence level.  This is our best fit model of the data, with a reduced $\chi ^2$  of 1.7.  This fit corresponds to a 2-10 keV flux of 4.7$^{+0.6}_{-1.1} \times 10^{-12}$ erg cm$^{-2}$ s$^{-1}$ (error denotes 90\% confidence interval).}
\end{figure}

\section{Luminosity of \emph{ESO 139-G12}}

The AGN \emph{ESO 139-G12} is correlated with two UHECRs, making it a particularly interesting source candidate.  This source is bright enough to determine its flux by fitting the spectrum using {\tt xspec}. The fit, assuming an absorbed power law, returned a hard photon index of 0.72$^{+0.09}_{-0.09}$ (error denotes 90\% confidence interval) and very low absorption, N$_{\rm H} < 8.2 \times 10^{21} {\rm cm}^{-2}$ at 90\% confidence level (Figure 1). From this fit we extracted a 2-10 keV flux of 4.7$^{+0.6}_{-1.1} \times 10^{-12}$ erg cm$^{-2}$ s$^{-1}$ (error denotes 90\% confidence interval).  When combined with its redshift, $z = 0.017$ ($\approx 74$ Mpc assuming a standard cosmology with $H_o = 70$), we find its 2-10 keV luminosity:  $L_{2-10} \approx 3.1^{+0.4}_{-0.7} \times 10^{42} \, {\rm erg  \  s^{-2}}$.  The conversion factor relating $L_{2-10}$ to the bolometric luminosity depends on the relative activity of the AGN \citep{LxConv}.  In the absence of a black hole mass measurement for this galaxy, we follow \citet{zfg09} and take $L_{\rm bol} = 20 \, L_{2-10}$ as is typical for an AGN with Eddington ratio less than 0.1.  This gives $L_{\rm bol} \approx 6.2 ^{+0.8}_{-1.4} \times 10^{43} \, $ erg s$^{-1}$ -- the first measurement of the bolometric luminosity of \emph{ESO 139-G12}.

Although \emph{ESO 139-G12} is the only galaxy we observed which appears to have nuclear activity, it is always possible that the host galaxy simply dominates the AGN energetically, making the AGN impossible to detect. Therefore the 2-10 keV luminosity is given for the other galaxies as well.   With an appropriate conversion factor, this can be used to get an upper limit on possible weak nuclear activity in the galaxy which may be of interest in other studies of UHECR sources.

\section{Conclusions}

These \emph{Chandra} observations complete the task of determining which galaxies found to correlate with UHECRs by \citet{augerScience07} and \citet{bzf10} have active nuclei, and of determining the bolometric luminosities of the AGNs correlating with UHECRs.
The X-ray fluxes of the ten unclassified correlating galaxies we studied all fall within the range typical of normal and starburst galaxies of the same optical magnitude, hence we find no evidence of AGN activity in any of them.\footnote{Of course, it is impossible to rule out the possibility of very low luminosity nuclear activity which is energetically dominated by the host galaxy, but such weakly active AGN are not well-motivated candidates for UHECR acceleration.}

Thus only 14 out of 27 UHECRs (13 out of 22 UHECRs with $|b| \geq 10^{\circ}$) correlate within $3.2^\circ$ with an actual AGN, using the Auger scan method to correlate UHECRs with VCV galaxies but discarding candidate sources which are not in fact AGNs.  Of the 13 highly luminous IR galaxies ($L_{\rm IR} \geq 10^{11.5} \, L_\odot$) found to correlate within $2.1^\circ$ with one (or more) of the 22 UHECRs with $|b| \geq 10^{\circ}$, we find that only five are also AGNs.\footnote{We could not run this correlation for the full 27 UHECRs because IRAS does not observe within the Galactic plane.}  Five of the 27 UHECRs, and one of the 22 UHECRs with $|b| \geq 10^{\circ}$, have neither an AGN nor a LIRG within 3.2$^\circ$ or $2.1^\circ$ respectively.   These uncorrelated UHECRs do not preferentially have low energies, for which sources are typically more distant, so the lack of correlation cannot be attributed entirely to incompleteness of the VCV or PSCz catalogs at larger redshifts.  

The degree of correlation with VCV galaxies was found to be lower in the second Auger data release \citep{augercollectedICRC09}.  It has also been shown recently that the scan method itself is subject to large sample-to-sample variance and that it systematically tends to overestimate the degree of correlation (Farrar et al., 2011, in preparation). These facts may suggest that other candidate sources not associated preferentially with AGNs may be responsible for some, most, or all UHECRs.  But other factors can contribute to a lack of correlation between UHECRs and AGNs as well, even if active galaxies are an important source of UHECRs:  incompleteness of source catalogs (especially within the Galactic plane), magnetic deflection that obliterates angular correlation between the CRs and their true source (particularly for UHECRs with charge $Z >1$), or transient production of UHECRs via exceptional, powerful flares flares in very weakly or non-active galaxies \citep{fg09}.

Another possibility raised by our results is that the VCV catalog may be so badly contaminated with non-AGNs, that the correlation analyses applied to the data up to now, c.f., \citet{augercollectedICRC09}, may be thrown off.  The ``spot check" of VCV provided by this work combined with \citep{zfg09} is the most thorough to date, by providing definitive identifications of AGN candidates;  if it gives a representative picture of the full VCV catalog, a third of the galaxies in the VCV catalog are not true AGNs.  Earlier spot-checks using (incomplete) data available in the literature, but considering a larger angular region around UHECRs and/or later UHECR catalogs, e.g., \citet{zbf10}, find that 1/4 - 1/3 of the AGN candidates in VCV are not AGNs.  The impact of a contamination of this magnitude on statistical correlation analyses such as Auger's scan method needs to be examined with simulations -- it may bias the results in unpredictable ways. 

The one established AGN which we observed in order to determine its bolometric luminosity, \emph{ESO 139-G12}, proves to have $L_{\rm bol}$ comparable to most of the other 13 correlated AGNs \citep{zfg09}.  Knowledge of $L_{\rm bol}$ and the energies of the correlated UHECRS (89 and 59 EeV) allows us to evaluate $\lambda_{\rm bol} \equiv 10^{-45} L_{\rm bol} \, E_{100}^{-2}$, the figure-of-merit introduced by \citet{zfg09} to quantify the ability of an AGN to accelerate a proton to the energy of the correlated UHECR.   A value of $\lambda_{\rm bol} \gtrsim 1$ satisfies the acceleration criterion for protons (c.f., \citet{fg09}).  With $\lambda_{\rm bol} \equiv 0.1$, \emph{ESO 139-G12} is weak for standard UHECR acceleration mechanisms.



\section{Summary}
Our observations and analysis using the \emph{Chandra} X-ray satellite and other data establish that one-third of the 21 galaxies in the Veron-Cetty Veron catalog of AGN candidates found by \citet{augerScience07} to correlate with UHECR arrival directions, do not in fact have active nuclei.  Combining this with our measurement of the X-ray luminosity of \emph{ESO 139-G12}, an AGN correlating with 2 UHECRs, implies that only one of the 27 UHECRs in that first Auger data-release correlates with an AGN that is powerful enough in its steady-state to accelerate protons to the observed energies, according to conventional acceleration mechanisms.  However 60\% of the UHECRs with $|b|>10^\circ$ do correlate with genuine but weak AGNs, consistent with the possibility that AGNs may have a flaring state that can accelerate UHECRs \citep{fg09}.  These results, along with the second Auger data release showing a lower correlation even with weak AGNs \citep{augercollectedICRC09}, underscore the fact that the source(s) of the highest energy cosmic rays,  as well as their composition, remains an open question.

This research was supported by NASA through Chandra Award No. GO9-0130X.   In addition, the research of GRF has been supported in part by NSF-PHY-0701451 and NSF-PHY-0900631.   GRF and IZ acknowledge their membership in the Pierre Auger Collaboration. 


\clearpage

\end{document}